\documentclass[reprint,showpacs,preprintnumbers,pre,superscriptaddress]{revtex4-1}
\usepackage[T1]{fontenc}
\usepackage[latin9]{inputenc}
\usepackage{amsmath}
\usepackage{amssymb}
\usepackage{graphicx}

\makeatletter
\usepackage{bm}\usepackage{amsfonts}
\usepackage{color}

\makeatother

\begin{document}
\title{Thermodynamic Geometric Control of Active Matter}
\author{Yating Wang}
\affiliation{School of Physics and Astronomy, Beijing Normal University, Beijing
100875, China}
\author{Enmai Lei}
\affiliation{School of Physics and Astronomy, Beijing Normal University, Beijing
100875, China}
\author{Yu-Han Ma}
\affiliation{School of Physics and Astronomy, Beijing Normal University, Beijing
100875, China}
\author{Z. C. Tu}
\email{tuzc@bnu.edu.cn}

\affiliation{School of Physics and Astronomy, Beijing Normal University, Beijing
100875, China}
\author{Geng Li}
\email{gengli@bnu.edu.cn}

\affiliation{School of Systems Science, Beijing Normal University, Beijing 100875,
China}
\begin{abstract}
Active matter represents a class of non-equilibrium systems that constantly
dissipate energy to produce directed motion. The thermodynamic control
of active matter holds great potential for advancements in synthetic
molecular motors, targeted drug delivery, and adaptive smart materials.
However, the inherently non-equilibrium nature of active matter poses
a significant challenge in achieving optimal control with minimal
energy cost. In this work, we extend the concept of thermodynamic
geometry, traditionally applied to passive systems, to active matter,
proposing a systematic geometric framework for minimizing energy cost
in non-equilibrium driving processes. We derive a cost metric that
defines a Riemannian manifold for control parameters, enabling the
use of powerful geometric tools to determine optimal control protocols.
The geometric perspective reveals that, unlike in passive systems,
minimizing energy cost in active systems involves a trade-off between
intrinsic and external dissipation, leading to an optimal transportation
speed that coincides with the self-propulsion speed of active matter.
This insight enriches the broader concept of thermodynamic geometry.
We demonstrate the application of this approach by optimizing the
performance of an active monothermal engine within this geometric
framework. 
\end{abstract}
\maketitle
\emph{Introduction.}--The thermodynamic control of active matter
is an emerging field that integrates principles of non-equilibrium
thermodynamics and statistical mechanics with the study of active
systems \cite{Fodor2022}. Active matter refers to systems that continuously
consume energy to generate self-sustained activity, leading to complex
collective behaviors \cite{Bechinger2016}. Understanding and controlling
these systems from a thermodynamic perspective holds immense potential
for a wide range of applications, including optimizing the performance
of synthetic molecular motors to mimic biological functions \cite{Palacci2013,Vutukuri2020},
designing low-energy drug delivery pathways to target specific sites
\cite{Ahmad2017,Iram2020,Ilker2022}, and developing smart materials
that can dynamically adapt to external stimuli, like temperature,
pressure, or electric fields \cite{Bril2022,Yildirim2023}. Achieving
these tasks often requires the rapid steering of active matter toward
a desired state. However, the inherently non-equilibrium nature of
active matter introduces a complex interplay between external control
and internal activity, complicating the prediction of the most efficient
protocols for driving the system.
\begin{figure}[!htp]
\includegraphics{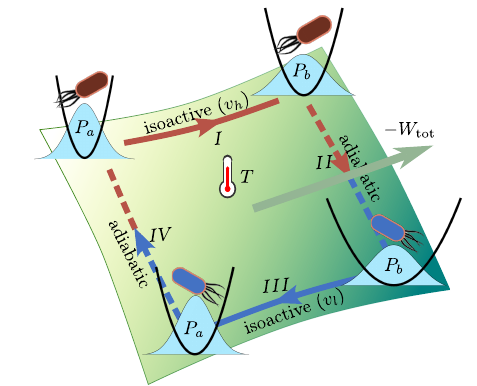} \caption{Scheme for an active monothermal engine with performance optimized
in a geometric space. The black curve represents a time-dependent
control potential $U(\mathbf{r},t)$ applied to the active matter,
while the filled blue curve indicates the boundary conditions $P_{a}$
and $P_{b}$ of the probability distribution $P(\mathbf{r},t)$. Steps
$I$ and $III$ correspond to isoactive processes with different activity
levels, $v_{h}$ and $v_{l}$, respectively. Steps $II$ and $IV$
are adiabatic processes where the activity switches instantaneously
between $v_{h}$ and $v_{l}$, while the system distribution holds
unchanged. Throughout the cyclic process, the active matter remains
in contact with a thermal bath at a constant temperature $T$ to produce
work $-W_{\mathrm{tot}}$, which can be optimized using the geometric
approach proposed in this work.}
\label{fig1}
\end{figure}

The quest for finding optimal control protocols with minimal energy
cost in passive systems is a critical problem that has been thoroughly
explored within the context of stochastic thermodynamics \cite{Seifert2012,Deffner2020,GueryOdelin2019,GueryOdelin2023}.
One of the most systematic approaches for addressing this is thermodynamic
geometry, which transforms the challenge of designing an optimal control
protocol into the problem of searching the geodesic path in a geometric
space defined by control parameters \cite{Crooks2007,Sivak2012,Li2022}.
As the controllable dimensions of the parametric space grow sufficiently
expressive, thermodynamic geometry converges with optimal transport
theory \cite{Benamou2000,Aurell2011,Dechant2019,VanVu2023}, which
maps a source distribution to a target distribution with minimal entropy
production \cite{Zhong2024}. This geometric approach has been widely
applied to various non-equilibrium systems, such as the Ising model
\cite{Gingrich2016,Louwerse2022}, bit initialization \cite{Ma2022,Boyd2022,Li2024},
and artificial thermal machines \cite{Abiuso2020,Zhang2023,Wang2024}.
The success of this approach motivates us to extend it to the search
for optimal control protocols beyond inherent passive systems. However,
how this geometric approach can be adapted to active matter remains
an open question.

In this work, we propose a systematic geometric framework for determining
the optimal control protocols with minimal energy cost in active systems.
We demonstrate that the energy cost of thermodynamic control is governed
by a positive cost metric, which can be minimized using optimal transport
theory. When the controllable parameters are limited, this cost metric
defines a Riemannian manifold spanned by the control parameters, and
the optimal control protocol is obtained through the thermodynamic
geometry scheme. In contrast to passive systems, where energy minimization
leads to a vanishing transportation speed, we find that minimizing
entropy production in active systems requires balancing intrinsic
and external dissipation, suggesting the existence of a finite optimal
transportation speed. The geometric viewpoint further shows that such
an optimal transportation speed just corresponds to the self-propulsion
speed of active matter. As shown in Fig.$\ $\ref{fig1}, we apply
this framework to optimize the performance of an active monothermal
engine, demonstrating the practicality of our approach in driving
active matter systems..

\textit{Theoretical model of active matter.}--Consider active matter
immersed in a thermal bath at a constant temperature $T$, controlled
by the potential $U(\mathbf{r},t)$, where $\mathbf{r}\equiv(r_{1},r_{2},r_{3})$
represents the three-dimensional spatial coordinate. The motion of
the active matter is governed by the Langevin equation: 
\begin{equation}
\gamma\dot{\mathbf{r}}=-\nabla U+\boldsymbol{\xi}+\gamma v\mathbf{n},\text{\ \ensuremath{\dot{\mathbf{n}}}=\ensuremath{\boldsymbol{\chi}}\ensuremath{\times}\ensuremath{\mathbf{n}},}\label{eq:withorent}
\end{equation}
where $v\mathbf{n}$ represents the self-propulsion velocity with
a constant norm $v$ and unit orientation vector $\mathbf{n}$. The
terms $\boldsymbol{\xi}$ and $\ensuremath{\boldsymbol{\chi}}$ are
uncorrelated Gaussian white noises with zero mean and variances $\langle\xi_{i}(t)\xi_{j}(t')\rangle=2\gamma T\delta_{ij}\delta(t-t')$
and $\langle\chi_{i}(t)\chi_{j}(t')\rangle=2(T/\gamma_{r})\delta_{ij}\delta(t-t')$,
respectively. Here, $\gamma$ and $\gamma_{r}$ represent the translational
and rotational friction coefficients. For simplicity, the Boltzmann
constant $k_{B}$ is set to unity. Considering a homogeneous active
system, we average over the rotational degrees of freedom to obtain
a theoretically tractable dynamical equation \cite{Fox1986,Fily2012,Farage2015},
\begin{equation}
\gamma\dot{\mathbf{r}}=-\nabla U+\boldsymbol{\xi}+\boldsymbol{\eta},\label{eq:noorent}
\end{equation}
where $\boldsymbol{\eta}$ is Gaussian colored noise with $\langle\eta_{i}(t)\rangle=0$
and $\langle\eta_{i}(t)\eta_{j}(t')\rangle=(\gamma TD_{a}/\tau_{p})\text{\ensuremath{\delta_{ij}}}\mathrm{e}^{-|t-t'|/\tau_{p}}$.
Here, $\tau_{p}\equiv\gamma_{r}/(2T)$ is the persistence time, and
$D_{a}\equiv\gamma v^{2}\tau_{p}/(3T)$ is the activity parameter. 

In a dilute suspension, where $\tau_{p}$ is much shorter than the
mean collision time, the evolution of the active matter's probability
distribution $P(\mathbf{r},t)$ can be described using the Fokker-Planck
equation:
\begin{equation}
\frac{\partial P(\mathbf{r},t)}{\partial t}=-\nabla\cdot\boldsymbol{J}(\mathbf{r},t)\label{eq:approx FFE}
\end{equation}
with the probability current $\boldsymbol{J}(\mathbf{r},t)\equiv-(1/\gamma)[(\nabla U)P+T(1+D_{a})\nabla P]$,
keeping terms up to the first order in the persistence time $\tau_{p}$.
Detailed derivations of Eq.$\text{\ }$(\ref{eq:approx FFE}) are
provided in Supplemental Material \cite{Supple}. 

\textit{Energy cost of thermodynamic control.}--The key challenge
in controlling active matter lies in finding an optimal control protocol
that minimizes the energy cost during a non-equilibrium process over
a time interval $[0,\tau]$. The energy cost is quantified by the
mean input work \cite{Jarzynski1997,Sekimoto1997}:
\begin{eqnarray}
W & \equiv & \int_{0}^{\tau}\langle\frac{\partial U}{\partial t}\rangle dt=\Delta\langle U\rangle-\int_{0}^{\tau}\langle\nabla U\circ\dot{\mathbf{r}}\rangle dt\nonumber \\
 & = & \Delta\langle U\rangle-T(1+D_{a})\Delta S+\int_{0}^{\tau}dt\int d\mathbf{r}\frac{\gamma\boldsymbol{J}^{2}}{P},\label{eq:meanwork}
\end{eqnarray}
where $\langle\cdot\rangle$ denotes an ensemble average over stochastic
trajectories with $\Delta\langle U\rangle\equiv\langle U\rangle|_{0}^{\tau}$,
and the circle $\circ$ between two stochastic variables represents
Stratonovich calculus. $S(t)\equiv-\int d\mathbf{r}P\ln P$ is the
Gibbs entropy with $\Delta S\equiv S(t)|_{0}^{\tau}$. Please see
Supplemental Material \cite{Supple} for detailed derivations of Eq.$\text{\ }$(\ref{eq:meanwork}).
The first two terms in Eq.$\ $(\ref{eq:meanwork}) are determined
by the given initial and final probability distributions. Thus, the
optimization focuses on the third term, $\gamma\int_{0}^{\tau}dt\int d\mathbf{r}\boldsymbol{J}^{2}/P$,
which defines a cost metric in the space of probability distributions
\cite{Benamou2000,Aurell2011,Dechant2019}. The optimal scheme of
transforming an active system from one probability distribution to
another can be solved by using optimal transport theory. Interestingly,
the irreversible part of the mean work for active matter is formally
equivalent to that of a passive Brownian particle system \cite{Seifert2005}.

Through a parametric design of the evolution path $P(\mathbf{r},t)=P(\mathbf{r},\boldsymbol{\lambda}(t))$
connecting the initial and final states, the potential can be decomposed
as $U=U_{o}+U_{a}$, where the original potential is defined as $U_{o}(\mathbf{r},\boldsymbol{\lambda})\equiv-T(1+D_{a})\ln P(\mathbf{r},\boldsymbol{\lambda})$,
and the auxiliary potential is $U_{a}\equiv U-U_{o}$. Here, $\boldsymbol{\lambda}(t)\equiv(\lambda_{1},\lambda_{2},\cdots,\lambda_{M})$
represents various time-dependent control parameters. It has been
shown that the auxiliary potential follows the form $U_{a}=\dot{\boldsymbol{\lambda}}\cdot\boldsymbol{f}(\mathbf{r},\boldsymbol{\lambda})$,
where $\boldsymbol{f}$ is determined by the evolution equation in
Eq.$\ $(\ref{eq:approx FFE}) \cite{Supple}. The irreversible part
of the mean work $W_{o}$ can then be expressed in a geometric form
\cite{Supple}:
\begin{eqnarray}
W_{o} & \equiv & W-\Delta\langle U\rangle+T(1+D_{a})\Delta S\nonumber \\
 & = & \sum_{\mu\nu}\int_{0}^{\tau}dt\dot{\lambda}_{\mu}\dot{\lambda}_{\nu}g_{\mu\nu}\label{eq:geowork}
\end{eqnarray}
where the positive semi-definite metric $g_{\mu\nu}\equiv(1/\gamma)\sum_{i}\int(\partial f_{\mu}/\partial r_{i})(\partial f_{\nu}/\partial r_{i})P(\mathbf{r},\boldsymbol{\lambda})d\mathbf{r}$
induces a Riemannian manifold on the parametric space \cite{Li2022}.
Thus, the challenge of minimizing energy cost in active systems can
be reinterpreted as finding the geodesic path in the parametric space.
This ``thermodynamic geometry'' scheme has been successfully applied
to passive systems for solving optimal control protocols \cite{Salamon1983,Crooks2007,Sivak2012,Li2023}.
Here, we extend this geometric framework to active matter, demonstrating
its adaptability for optimizing thermodynamic control in persistently
far-from-equilibrium conditions. This is our first main result.

\textit{Optimal control duration.}--Based on the framework of stochastic
thermodynamics \cite{Sekimoto2010,Seifert2012}, we define the mean
absorbed heat during the control of active matter as:
\begin{eqnarray}
Q & \equiv & \int_{0}^{\tau}\langle(-\gamma\dot{\mathbf{r}}+\boldsymbol{\xi})\circ\dot{\mathbf{r}}\rangle dt=\int_{0}^{\tau}\langle(\nabla U-\boldsymbol{\eta})\circ\dot{\mathbf{r}}\rangle dt\nonumber \\
 & = & T\Delta S-\gamma v^{2}\tau-\int_{0}^{\tau}dt\int d\mathbf{r}[\frac{\gamma\boldsymbol{J}^{2}}{P}-\frac{T\tau_{p}v^{2}(\nabla P)^{2}}{3P}]\label{eq:meanheat}
\end{eqnarray}
with the derivations of the third equality presented in Supplemental
Material \cite{Supple}. To facilitate comparison with passive systems,
we rewrite Eq.$\ $(\ref{eq:meanheat}) as a balance equation for
the entropy production of active matter:
\begin{eqnarray}
\Delta S_{\mathrm{tot}} & \equiv & \Delta S-\frac{Q}{T}\nonumber \\
 & = & \frac{\gamma v^{2}\tau}{T}+\int_{0}^{\tau}dt\int d\mathbf{r}[\frac{\gamma\boldsymbol{J}^{2}}{TP}-\frac{\tau_{p}v^{2}(\nabla P)^{2}}{3P}],\label{eq:entropyproduc}
\end{eqnarray}
which is always positive since the last term is a first order small
quantity of $\tau_{p}$. The first term in Eq.$\ $(\ref{eq:entropyproduc})
arises from sustained energy dissipation to maintain the activity,
which increases linearly with the control duration $\tau$. The second
term represents entropy production due to the non-equilibrium control,
which takes the same form as in passive systems \cite{Seifert2005}.
Unlike in passive systems, the results in Eq.$\ $(\ref{eq:entropyproduc})
can be treated as a modified second law of thermodynamics in active
systems.

When considering an optimal control protocol, solved using either
optimal transport theory \cite{Dechant2019} or the thermodynamic
geometry scheme \cite{Li2022}, we obtain the scaling relations $P(\mathbf{r},t)=\mathcal{P}(\mathbf{r},s)$
and $\boldsymbol{J}(\mathbf{r},t)=\mathcal{\boldsymbol{J}}(\mathbf{r},s)/\tau$
with normalized time $s\equiv t/\tau$, as proven in Supplemental
Material \cite{Supple}. The entropy production in Eq.$\ $(\ref{eq:entropyproduc})
can then be rewritten as a scaling relation:
\begin{eqnarray}
\Delta S_{\mathrm{tot}} & = & \frac{\gamma v^{2}\tau}{T}+\int_{0}^{1}ds\int d\mathbf{r}[\frac{\gamma\mathcal{\boldsymbol{J}}^{2}}{T\mathcal{P}\tau}-\frac{\tau_{p}v^{2}(\nabla\mathcal{P})^{2}\tau}{3\mathcal{P}}]\nonumber \\
 & = & \frac{E_{a}}{T}\tau+\frac{A_{o}}{T}\frac{1}{\tau},\label{eq:scalingentropyprod}
\end{eqnarray}
where $E_{a}\equiv\gamma v^{2}[1-\tau_{p}\int_{0}^{1}ds\int d\mathbf{r}T(\nabla\mathcal{P})^{2}/(3\gamma\mathcal{P})]$
and $A_{o}\equiv\gamma\int_{0}^{1}ds\int d\mathbf{r}\mathcal{\boldsymbol{J}}^{2}/\mathcal{P}$.
The first term in the second line of Eq.$\ $(\ref{eq:scalingentropyprod})
reveals that the non-equilibrium dissipation from the activity accumulates
linearly with the control time. The second term indicates that the
irreversible energy cost of the optimal control follows a $1/\tau$
scaling, a behavior widely discussed in passive systems \cite{Broeck2005,Schmiedl2007,Esposito2010,Ryabov2016,Ma2020}.
The observation in Eq.$\ $(\ref{eq:scalingentropyprod}) aligns with
the results obtained by Davis \textit{et al.}, who found a similar
scaling relation for the optimal control of active systems in the
slow and weak driving limit \cite{Davis2024}. Here we demonstrate
that this scaling relation applies to optimal thermodynamic control
of active systems with arbitrary driving rates. This is our second
major result.

This scaling relation implies an optimal control duration $\tau^{*}=\sqrt{A_{o}/E_{a}}$
for achieving minimum entropy production, $\Delta S_{\mathrm{tot}}^{\mathrm{min}}=(2/T)\sqrt{A_{o}E_{a}}$,
in an active system. This contrasts with the monotonically decreasing
relationship between entropy production and control duration found
in passive systems. In the geometric space, the minimum $A_{o}$ can
be expressed as the square of the Wasserstein distance \cite{Benamou2000},
$\mathrm{min}\text{\ }A_{o}=\gamma\mathcal{W}^{2}$, where the Wasserstein
distance is defined as $\mathcal{W}\equiv\int_{0}^{1}ds\sqrt{\int d\mathbf{r}\mathcal{\boldsymbol{J}}^{2}/\mathcal{P}}$
describing the shortest distance between the initial and final states.
The optimal control duration with minimum entropy production follows
as $\tau^{*}=\mathcal{W}/[v\sqrt{1-\tau_{p}\int_{0}^{1}ds\int d\mathbf{r}T(\nabla\mathcal{P})^{2}/(3\gamma\mathcal{P})}]$.
Retaining terms up to the zero order in the persistence time $\tau_{p}$,
we obtain that $E_{a}\approx\gamma v^{2}$ and $A_{o}\approx\gamma\int_{0}^{1}ds\int d\mathbf{r}\mathcal{\boldsymbol{J}}_{0}^{2}/\mathcal{P}$,
where $\boldsymbol{\mathcal{J}}_{0}(\mathbf{r},t)\equiv-(1/\gamma)[(\nabla U)\mathcal{P}+T\nabla\mathcal{P}]$.
The minimum entropy production is determined by the Wasserstein distance
$\mathcal{W}\approx\int_{0}^{1}ds\sqrt{\int d\mathbf{r}\mathcal{\boldsymbol{J}}_{0}^{2}/\mathcal{P}}$
with the optimal control duration obtained as $\tau^{*}=\mathcal{W}/v$.

In passive systems, the geometric viewpoint suggests that the shortest
path for transforming one state to another with minimum entropy production
follows the geodesic line, whose length corresponds to the Wasserstein
distance. The optimal transportation speed along the geodesic line
is constant and approaches zero as we aim for minimum entropy production
\cite{Benamou2000}. In sharp contrast, in active systems, the optimal
transportation speed along the geodesic line has a finite value $v$,
which happens to be the self-propulsion speed of the active system.
When the speed is lower than $v$, dissipation from intrinsic activity
dominates, whereas at speeds higher than $v$, energy cost from the
external control becomes the dominant factor. These geometric perspectives
lead to our third main result: our geometric framework not only provides
a systematic approach for optimizing the thermodynamic control of
active matter, but also imparts a clear geometric meaning to self-propulsion
in active matter, enriching the broader concept of thermodynamic geometry.

\textit{Active monothermal engines.}--As a practical application,
we construct a cyclic engine operating with active matter to extract
useful work and systematically investigate the optimization of its
performance. As shown in Fig.$\ $\ref{fig1}, the active engine consists
of two isoactive processes and two adiabatic processes with inverse
boundary conditions $P_{a}(\mathbf{r})$ and $P_{b}(\mathbf{r})$.
Unlike passive thermal engines that operate between baths at different
temperatures, the two isoactive processes in active engines operate
within a single bath but with different self-propulsion speeds of
active matter, where $v_{h}>v_{l}$. During the adiabatic process,
the self-propulsion speed instantaneously changes between $v_{h}$
and $v_{l}$ while the system distribution keeps unaltered. Several
experiments have confirmed the possibility of modulating the activity
based on various physical or chemical factors \cite{Buttinoni2012,Krishnamurthy2016,Vutukuri2020,Militaru2021,Baldovin2023}.

\begin{figure}[!htp]
\includegraphics{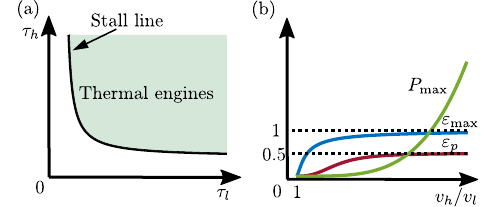} \caption{Scheme for the performance of an active monothermal engine. (a) The
stall line $A_{o}/\tau_{h}+A_{o}/\tau_{l}=T(D_{a}^{h}-D_{a}^{l})\Delta S$
for the active monothermal engine. Above this line, the cyclic process
operates as a thermal engine, producing work. Below this line, the
active cyclic process requires work input to function. (b) The maximum
efficiency $\varepsilon_{\mathrm{max}}$ (blue curve), maximum power
$\mathrm{P}_{\mathrm{max}}$ (green curve), and the efficiency at
maximum power $\varepsilon_{p}$ (red curve) are shown as functions
of the ratio between activity levels $v_{h}/v_{l}$. Here, the lower
activity $v_{l}$ is fixed. Both the maximum efficiency $\varepsilon_{\mathrm{max}}$,
maximum power $\mathrm{P}_{\mathrm{max}}$, and the efficiency at
maximum power $\varepsilon_{p}$ increase with the ratio $v_{h}/v_{l}$.
As the ratio $v_{h}/v_{l}$ grows, $\varepsilon_{\mathrm{max}}$ approaches
$1$, while $\varepsilon_{p}$ approaches $1/2$.}
\label{fig2}
\end{figure}

In a cyclic process, there is no net change in the potential energy
or entropy of active systems, such that $\sum_{i=1}^{4}\Delta\langle U\rangle^{(i)}=\sum_{i=1}^{4}\Delta S^{(i)}=0$.
Since entropy remains constant during an adiabatic process, the entropy
change in the isoactive process is $\Delta S^{(1)}=-\Delta S^{(3)}\equiv\Delta S$.
By applying the optimal control protocol derived from the geometric
approach to the isoactive process, the total output work and heat
during a cycle follow as:
\begin{eqnarray}
-W_{\mathrm{tot}} & = & T(D_{a}^{h}-D_{a}^{l})\Delta S-\frac{A_{o}}{\tau_{h}}-\frac{A_{o}}{\tau_{l}},\nonumber \\
-Q_{\mathrm{tot}} & = & \frac{A_{o}}{\tau_{h}}+\frac{A_{o}}{\tau_{l}}+E_{a}^{h}\tau_{h}+E_{a}^{l}\tau_{l},\label{eq:totalwork}
\end{eqnarray}
where $D_{a}^{h}$ and $D_{a}^{l}$ are the activity parameters, $E_{a}^{h}$
and $E_{a}^{l}$ are the scaling partial heat, and $\tau_{h}$ and
$\tau_{l}$ denote the durations of the isoactive processes with self-propulsion
speeds $v_{h}$ and $v_{l}$, respectively. The scaling irreversible
work $A_{o}=\gamma\mathcal{W}^{2}$ is uniform for both isoactive
processes due to the symmetry of the Wasserstein distance $\mathcal{W}(P_{a},P_{b})=\mathcal{W}(P_{b},P_{a})$.

The performance of the active monothermal engine can be evaluated
by investigating the power and efficiency during a finite-time cycle.
The efficiency $\varepsilon$ is defined as the ratio of the output
work to the total energy cost \cite{Hill1974,Juelicher1997,Pietzonka2016,Szamel2020,Fodor2021}:
\begin{eqnarray}
\varepsilon & \equiv & \frac{W_{\mathrm{tot}}}{W_{\mathrm{tot}}+Q_{\mathrm{tot}}}\nonumber \\
 & = & \frac{T(D_{a}^{h}-D_{a}^{l})\Delta S-A_{o}/\tau_{h}-A_{o}/\tau_{l}}{T(D_{a}^{h}-D_{a}^{l})\Delta S+E_{a}^{h}\tau_{h}+E_{a}^{l}\tau_{l}},\label{eq:efficiency}
\end{eqnarray}
which is always less than one. As shown in Fig.$\ $\ref{fig2}(a),
there exists a ``stall line'' $A_{o}/\tau_{h}+A_{o}/\tau_{l}=T(D_{a}^{h}-D_{a}^{l})\Delta S$,
where the engine stops producing work. Below this line, the engine
starts consuming work. The maximum efficiency is given by $\varepsilon_{\mathrm{max}}=(\tau_{h}^{*}/\tau_{h}^{e})^{2}=(\tau_{l}^{*}/\tau_{l}^{e})^{2}$,
where $\tau_{h}^{e}$ and $\tau_{l}^{e}$ represent the optimal driving
durations for maximum efficiency, and $\tau_{h}^{*}\equiv\sqrt{A_{o}/E_{a}^{h}}$
and $\tau_{l}^{*}\equiv\sqrt{A_{o}/E_{a}^{l}}$ represent the optimal
durations for minimum entropy production in the two isoactive processes.
This behavior differs from that of passive thermal engines, which
reach maximum efficiency at the long time limit. In Fig.$\ $\ref{fig2}(b),
$\varepsilon_{\mathrm{max}}$ is plotted as a function of the activity
ratio $v_{h}/v_{l}$. Here, the lower activity level $v_{l}$ remains
fixed while $v_{h}/v_{l}$ is varied. As the activity ratio increases,
$\varepsilon_{\mathrm{max}}$ rises and eventually approaches one.
When $v_{h}/v_{l}\to1$, the maximum efficiency scales as $\varepsilon_{\mathrm{max}}\propto(v_{h}/v_{l}-1)^{2}$,
while at the large ratio limit $v_{h}/v_{l}\to\infty$, the maximum
efficiency yields $\varepsilon_{\mathrm{max}}=1-\alpha v_{h}/v_{l}$,
where $\alpha$ is a constant. Detailed expressions for $\varepsilon_{\mathrm{max}}$,
$\tau_{h}^{e}$, $\tau_{l}^{e}$, $\alpha$, and the analytical relationship
between $\varepsilon_{\mathrm{max}}$ and $v_{h}/v_{l}$ are provided
in Supplemental Material \cite{Supple}.

The power of the active engine $\mathrm{P}\equiv-W_{\mathrm{tot}}/(\tau_{h}+\tau_{l})$
reaches its maximum: 
\begin{equation}
\mathrm{P}_{\mathrm{max}}=\frac{T^{2}(D_{a}^{h}-D_{a}^{l})^{2}(\Delta S)^{2}}{16A_{o}}\label{eq:maxpower}
\end{equation}
with optimal driving durations $\tau_{h}^{p}=\tau_{l}^{p}=4A_{o}/[T(D_{a}^{h}-D_{a}^{l})\Delta S]$.
In Fig.$\ $\ref{fig2}(b), the maximum power $\mathrm{P}_{\mathrm{max}}$
is plotted as a function of the activity ratio $v_{h}/v_{l}$, following
the analytical relationship $\mathrm{P}_{\mathrm{max}}\propto[(v_{h}/v_{l})^{2}-1]^{2}$
\cite{Supple}. The efficiency at maximum power is then obtained as
\begin{eqnarray}
\varepsilon_{p} & = & \frac{1}{2\{1+A_{o}(E_{a}^{h}+E_{a}^{l})/[T^{2}(D_{a}^{h}-D_{a}^{l})^{2}(\Delta S)^{2}]\}}\nonumber \\
 & = & \frac{1}{2[1+(E_{a}^{h}+E_{a}^{l})/(16\mathrm{P}_{\mathrm{max}})]},\label{eq:effmaxpow}
\end{eqnarray}
which approaches $1/2$ in the large ratio limit $v_{h}/v_{l}\to\infty$.
The relationship between $\varepsilon_{p}$ and $v_{h}/v_{l}$ is
plotted in Fig.$\ $\ref{fig2}(b), with the analytical relation provided
in Supplemental Material \cite{Supple}.

\emph{Conclusions.}--In summary, we have proposed a geometric approach
to optimize the thermodynamic control of active matter during finite-time
state transitions. In a dilute suspension, the irreversible energy
cost of active matter is characterized by a positive cost metric,
defined on the space of probability distributions. Optimal control
protocols are achieved by applying optimal transport theory. Considering
the limited controllable parametric space, which is more practical
for experiments and simulations, this cost metric describes a Riemannian
manifold spanned by the control parameters, with the optimal control
protocol corresponding to the geodesic path. Moreover, we have identified
a trade-off between intrinsic and external dissipation when minimizing
entropy production, revealing a constant optimal transportation speed
along the geodesic line for non-equilibrium driving processes of active
matter. The geometric perspective further suggests that the speed
happens to be the self-propulsion speed of the active system, which
endows the self-propulsion speed with a clear geometric significance.
This behavior contrasts with passive systems, where the optimal transportation
speed for minimum entropy production infinitely approaches zero. We
have also demonstrated the practical utility of our approach by optimizing
the performance of an active monothermal engine, showing that both
the maximum efficiency and maximum power monotonically increase with
the activity ratio. 

The geometric insights not only deepen our thermodynamic understanding
of active matter's control, but also provide geometric interpretation
of active matter's motion. The geometric framework establishes a direct
link between the minimal-energy-cost path for active matter and the
geodesic line in parametric space. Given a target state, optimal control
protocols can be systematically obtained by solving the geodesic equation
determined by the cost metric in Eq.$\ $(\ref{eq:geowork}), without
the need to individually analyze complex active matter systems. It
is intriguing to compare our theoretical findings on optimal control
of active matter, such as the scaling relation in Eq.$\ $(\ref{eq:scalingentropyprod}),
with biological evolution process in cells. Additionally, monothermal
engines have been realized experimentally with a passive particle
immersed in active bacterial baths \cite{Krishnamurthy2016}. It would
be valuable to experimentally compare the performance of our active
monothermal engine with that of the engine described in Ref.$\ $\cite{Krishnamurthy2016}.

\emph{Acknowledgement.}--This work is supported by the National Natural
Science Foundation of China (NSFC) (Grants No. 12405031, No. 12475032,
and No. 12305037 ) and the Fundamental Research Funds for the Central
Universities (Grant No. 2233100001).

\bibliographystyle{apsrev4-1}
\bibliography{ref}

\end{document}


\title{Supplementary Material: Thermodynamic Geometric Control of Active
Matter}
\author{Yating Wang}
\affiliation{School of Physics and Astronomy, Beijing Normal University, Beijing,
100875, China}
\author{Enmai Lei}
\affiliation{School of Physics and Astronomy, Beijing Normal University, Beijing,
100875, China}
\author{Yu-Han Ma}
\affiliation{School of Physics and Astronomy, Beijing Normal University, Beijing,
100875, China}
\author{Z. C. Tu}
\email{tuzc@bnu.edu.cn}

\affiliation{School of Physics and Astronomy, Beijing Normal University, Beijing,
100875, China}
\author{Geng Li}
\email{gengli@bnu.edu.cn}

\affiliation{School of Systems Science, Beijing Normal University, Beijing 100875,
China}

\maketitle
The supplementary materials are devoted to provide detailed derivations
in the main context.

\tableofcontents{}

\section{The Fokker-Planck equation for active matter\label{SSec-one}}

The dynamics of a homogeneous active system is governed by a non-Markovian
Langevin equation:
\begin{equation}
\gamma\dot{\mathbf{r}}=-\nabla U+\boldsymbol{\xi}+\boldsymbol{\eta}.\label{seq:lanequcolored}
\end{equation}
Applying the Fox approximation \cite{Fox1986,Fox1986a,Farage2015},
the evolution equation for the probability distribution $P(\mathbf{r},t)$
is given by
\begin{equation}
\frac{\partial P(\mathbf{r},t)}{\partial t}=-\nabla\cdot\boldsymbol{J}(\mathbf{r},t),\label{seq:foxfpequ}
\end{equation}
where the probability current is:
\begin{equation}
\boldsymbol{J}(\mathbf{r},t)\equiv-\frac{1}{\gamma}[(\nabla U(\mathbf{r},t))P(\mathbf{r},t)+T\nabla(\alpha(\mathbf{r},t)P(\mathbf{r},t))].\label{seq:foxcurrent}
\end{equation}
The dimensionless parameter $\alpha(\mathbf{r},t)$ is defined as:
\begin{equation}
\alpha(\mathbf{r},t)\equiv1+\frac{D_{a}}{1+\frac{\tau_{p}}{\gamma}\nabla^{2}U(\mathbf{r},t)}.\label{seq:dimparat}
\end{equation}
 In a dilute suspension, where the persistence time $\tau_{p}$ is
much smaller than the mean collision time, keeping terms up to the
first order in $\tau_{p}$, the dimensionless parameter simplifies
to $\alpha\approx1+D_{a}$. Therefore, the probability current reduces
to:
\begin{equation}
\boldsymbol{J}(\mathbf{r},t)\approx-\frac{1}{\gamma}[(\nabla U(\mathbf{r},t))P(\mathbf{r},t)+T(1+D_{a})\nabla P(\mathbf{r},t)],\label{seq:smallcurrent}
\end{equation}
which is equivalent to the current in Eq.$\ $(3) of the main text.
According to the probability current in Eq.$\text{\ }$(\ref{seq:smallcurrent}),
the equivalent Langevin equation becomes:
\[
\gamma\dot{\mathbf{r}}=-\nabla U+\boldsymbol{\xi}+\boldsymbol{\xi}_{a},
\]
where $\boldsymbol{\xi}_{a}$ represents Gaussian white noise with
zero mean and variances $\langle\xi_{a}^{(i)}(t)\xi_{a}^{(j)}(t')\rangle=2\gamma TD_{a}\delta_{ij}\delta(t-t')$.

\section{The mean input work done for active matter\label{SSec-two}}

Based on the framework of stochastic thermodynamics \cite{Jarzynski1997,Sekimoto1997},
the mean work performed in a non-equilibrium driving process is expressed
as:
\begin{eqnarray}
W & \equiv & \int_{0}^{\tau}\langle\frac{\partial U}{\partial t}\rangle dt=\Delta\langle U\rangle-\int_{0}^{\tau}\langle\nabla U\circ\dot{\mathbf{r}}\rangle dt,\label{seq:intepart}
\end{eqnarray}
where we have employed integration by parts. To evaluate the ensemble
average in the second term, we apply a path integral formulation:
\begin{eqnarray}
\langle\nabla U\circ\dot{\mathbf{r}}\rangle & = & \int D[\mathbf{r}(t)]\mathcal{F}[\mathbf{r}(t)]\nabla U(\mathbf{r}(t),t)\circ\dot{\mathbf{r}}(t)\nonumber \\
 & = & \frac{1}{\gamma}\int D[\mathbf{r}(t)]\mathcal{F}[\mathbf{r}(t)]\nabla U(\mathbf{r}(t),t)\circ[-\nabla U(\mathbf{r}(t),t)+\boldsymbol{\xi}(t)+\boldsymbol{\xi}_{a}(t)]\nonumber \\
 & = & \frac{1}{\gamma}\int D[\mathbf{r}(t)]\mathcal{F}[\mathbf{r}(t)]\int d\mathbf{r}\delta(\mathbf{r}-\mathbf{r}(t))\nabla U(\mathbf{r}(t),t)\circ[-\nabla U(\mathbf{r}(t),t)+\boldsymbol{\xi}(t)+\boldsymbol{\xi}_{a}(t)]\nonumber \\
 & = & -\frac{1}{\gamma}\int d\mathbf{r}[\nabla U(\mathbf{r},t)]^{2}P(\mathbf{r},t)+\frac{1}{\gamma}\int d\mathbf{r}\nabla U(\mathbf{r},t)\circ[\langle\delta(\mathbf{r}-\mathbf{r}(t))\boldsymbol{\xi}(t)\rangle+\langle\delta(\mathbf{r}-\mathbf{r}(t))\boldsymbol{\xi}_{a}(t)\rangle].\label{seq:pathinteg}
\end{eqnarray}
In the fourth equality, we have employed the formulation $P(\mathbf{r},t)=\int D[\mathbf{r}(t)]\mathcal{F}[\mathbf{r}(t)]\delta(\mathbf{r}-\mathbf{r}(t))$,
where $\mathcal{F}[\mathbf{r}(t)]$ represents the path probability.
Considering the commutation between the noise terms $\boldsymbol{\xi}(t)$
and $\boldsymbol{\xi}_{a}(t)$ with the coordinate $\mathbf{r}$ \cite{Reichl1998,Li2022},
we derive:
\begin{equation}
\langle\delta(\mathbf{r}-\mathbf{r}(t))\boldsymbol{\xi}(t)\rangle=-T\nabla P(\mathbf{r},t)\label{seq:orignoise}
\end{equation}
and
\begin{equation}
\langle\delta(\mathbf{r}-\mathbf{r}(t))\boldsymbol{\xi}_{a}(t)\rangle=-TD_{a}\nabla P(\mathbf{r},t).\label{seq:activenoise}
\end{equation}
Substituting these into Eq.$\ $(\ref{seq:pathinteg}), we obtain:
\begin{eqnarray}
\langle\nabla U\circ\dot{\mathbf{r}}\rangle & = & -\frac{1}{\gamma}\int d\mathbf{r}\nabla U(\mathbf{r},t)\circ[(\nabla U(\mathbf{r},t))P(\mathbf{r},t)+T(1+D_{a})\nabla P(\mathbf{r},t)]\nonumber \\
 & = & \int d\mathbf{r}\nabla U(\mathbf{r},t)\circ\boldsymbol{J}(\mathbf{r},t)\nonumber \\
 & = & -\int d\mathbf{r}\frac{\gamma\boldsymbol{J}(\mathbf{r},t)+T(1+D_{a})\nabla P(\mathbf{r},t)}{P(\mathbf{r},t)}\circ\boldsymbol{J}(\mathbf{r},t)\nonumber \\
 & = & -\int d\mathbf{r}\frac{\gamma\boldsymbol{J}^{2}}{P}-T(1+D_{a})\int d\mathbf{r}\frac{\nabla P\circ\boldsymbol{J}}{P}.\label{seq:pocurrent}
\end{eqnarray}
In the third equality, we have used the definition of the probability
current in Eq.$\ $(\ref{seq:smallcurrent}). Considering the time
derivative of the Gibbs entropy $S(t)\equiv-\int d\mathbf{r}P\ln P$,
we find:
\begin{eqnarray}
\dot{S}(t) & = & -\int d\mathbf{r}\frac{\partial P}{\partial t}\ln P\nonumber \\
 & = & \int d\mathbf{r}\nabla\cdot\boldsymbol{J}(\mathbf{r},t)\ln P\nonumber \\
 & = & -\int d\mathbf{r}\frac{\nabla P\circ\boldsymbol{J}}{P}\label{seq:dotentropy}
\end{eqnarray}
By leveraging the relations in Eqs.$\ $(\ref{seq:pocurrent}) and$\ $(\ref{seq:dotentropy}),
we ultimately derive the mean work as:
\begin{equation}
W=\Delta\langle U\rangle-T(1+D_{a})\Delta S+\int_{0}^{\tau}dt\int d\mathbf{r}\frac{\gamma\boldsymbol{J}^{2}}{P},\label{seq:finalwork}
\end{equation}
which is equivalent to the form in Eq.$\ $(4) of the main text.

\section{The general form of the auxiliary potential\label{SSec-three}}

Given the initial and final states, we can design an evolution path
$P(\mathbf{r},t)=P(\mathbf{r},\boldsymbol{\lambda}(t))$ with the
time-dependent control parameters $\boldsymbol{\lambda}(t)\equiv(\lambda_{1},\lambda_{2},\cdots,\lambda_{M})$.
With this setting, we split the potential $U(\mathbf{r},t)$ as $U=U_{o}+U_{a}$
with the original potential defined as $U_{o}(\mathbf{r},\boldsymbol{\lambda})\equiv-T(1+D_{a})\ln P(\mathbf{r},\boldsymbol{\lambda})+F(\boldsymbol{\lambda})$.
Here, $F(\boldsymbol{\lambda})$ represents a normalization constant.
Substituting this splitting into Eq.$\ $(\ref{seq:foxfpequ}), we
get
\begin{equation}
\boldsymbol{\dot{\lambda}}(\nabla_{\boldsymbol{\lambda}}F-\nabla_{\boldsymbol{\lambda}}U_{o})=\frac{T(1+D_{a})}{\gamma}\nabla^{2}U_{a}-\frac{1}{\gamma}\nabla U_{o}\nabla U_{a}.\label{seq:uaequation}
\end{equation}
By comparing both sides of Eq.$\ $(\ref{seq:uaequation}), we arrive
at the general form of the auxiliary potential:
\begin{equation}
U_{a}(\mathbf{r},t)=\boldsymbol{\dot{\lambda}}\cdot\boldsymbol{f}(\mathbf{r},\boldsymbol{\lambda}),\label{seq:generaaxu}
\end{equation}
where the function $\boldsymbol{f}(\mathbf{r},\boldsymbol{\lambda})$
is determined by the evolution equation:
\begin{equation}
\nabla_{\boldsymbol{\lambda}}F-\nabla_{\boldsymbol{\lambda}}U_{o}=\frac{T(1+D_{a})}{\gamma}\nabla^{2}\cdot\boldsymbol{f}-\frac{1}{\gamma}\nabla U_{o}\nabla\cdot\boldsymbol{f}.\label{seq:auxdeter}
\end{equation}
With the general form of the auxiliary potential in Eq.$\ $(\ref{seq:generaaxu}),
the irreversible part of the mean work can be rewritten as:

\begin{eqnarray}
W_{o} & \equiv & W-\Delta\langle U\rangle+T(1+D_{a})\Delta S\nonumber \\
 & = & \int_{0}^{\tau}dt\int d\mathbf{r}\frac{\gamma\boldsymbol{J}^{2}}{P}\nonumber \\
 & = & \frac{1}{\gamma}\sum_{\mu\nu}\int_{0}^{\tau}dt\dot{\lambda}_{\mu}\dot{\lambda}_{\nu}\sum_{i}\int(\partial f_{\mu}/\partial r_{i})(\partial f_{\nu}/\partial r_{i})P(\mathbf{r},\boldsymbol{\lambda})d\mathbf{r}\nonumber \\
 & = & \sum_{\mu\nu}\int_{0}^{\tau}dt\dot{\lambda}_{\mu}\dot{\lambda}_{\nu}g_{\mu\nu}\label{seq:irrpartwork}
\end{eqnarray}
with the positive semi-definite metric $g_{\mu\nu}\equiv(1/\gamma)\sum_{i}\int(\partial f_{\mu}/\partial r_{i})(\partial f_{\nu}/\partial r_{i})P(\mathbf{r},\boldsymbol{\lambda})d\mathbf{r}$,
representing a geometric space spanned by control parameters. The
result in Eq.$\ $(\ref{seq:irrpartwork}) is equivalent to the geometric
expression in Eq.$\ $(5) of the main text.

\section{The mean absorbed heat during the control of active matter\label{SSec-four}}

The mean heat absorbed during non-equilibrium control of active matter
is expressed as:
\begin{eqnarray}
Q & \equiv & \int_{0}^{\tau}\langle(-\gamma\dot{\mathbf{r}}+\boldsymbol{\xi})\circ\dot{\mathbf{r}}\rangle dt=\int_{0}^{\tau}\langle(\nabla U(\mathbf{r}(t),t)-\boldsymbol{\xi}_{a})\circ\dot{\mathbf{r}}\rangle dt\nonumber \\
 & = & \int_{0}^{\tau}\langle\nabla U(\mathbf{r}(t),t)\circ\dot{\mathbf{r}}\rangle dt-\int_{0}^{\tau}\langle\boldsymbol{\xi}_{a}\circ\dot{\mathbf{r}}\rangle dt.\label{seq:meanabsorbed}
\end{eqnarray}
The first term is derived in Eq.$\ $(\ref{seq:pocurrent}), while
the second term:
\begin{eqnarray}
\int_{0}^{\tau}\langle\boldsymbol{\xi}_{a}\circ\dot{\mathbf{r}}\rangle dt & = & \int_{0}^{\tau}dt\langle\boldsymbol{\xi}_{a}(t)\circ\frac{1}{\gamma}(-\nabla U(\mathbf{r}(t),t)+\boldsymbol{\xi}(t)+\boldsymbol{\xi}_{a}(t))\rangle\nonumber \\
 & = & \frac{1}{\gamma}\int_{0}^{\tau}dt[-\langle\boldsymbol{\xi}_{a}(t)\circ\nabla U(\mathbf{r}(t),t)\rangle+\langle\boldsymbol{\xi}_{a}(t)\circ\boldsymbol{\xi}(t)\rangle+\langle\boldsymbol{\xi}_{a}(t)\circ\boldsymbol{\xi}_{a}(t)\rangle]\nonumber \\
 & = & -\frac{1}{\gamma}\int_{0}^{\tau}dt\int d\mathbf{r}\nabla U(\mathbf{r},t)\circ\langle\delta(\mathbf{r}-\mathbf{r}(t))\boldsymbol{\xi}_{a}(t)\rangle+\frac{1}{\gamma}\frac{3\gamma TD_{a}\tau}{\tau_{p}}\nonumber \\
 & = & -\frac{1}{\gamma}\int_{0}^{\tau}dt\int d\mathbf{r}\nabla U(\mathbf{r},t)\circ(-TD_{a}\nabla P(\mathbf{r},t))+\frac{3TD_{a}\tau}{\tau_{p}}\nonumber \\
 & = & \frac{TD_{a}}{\gamma}\int_{0}^{\tau}dt\int d\mathbf{r}\nabla U(\mathbf{r},t)\circ\nabla P(\mathbf{r},t)+\frac{3TD_{a}\tau}{\tau_{p}}\nonumber \\
 & = & -\frac{TD_{a}}{\gamma}\int_{0}^{\tau}dt\int d\mathbf{r}\frac{\gamma\boldsymbol{J}(\mathbf{r},t)+T(1+D_{a})\nabla P(\mathbf{r},t)}{P(\mathbf{r},t)}\circ\nabla P(\mathbf{r},t)+\frac{3TD_{a}\tau}{\tau_{p}}\nonumber \\
 & = & -TD_{a}\int_{0}^{\tau}dt\int d\mathbf{r}\frac{\nabla P\circ\boldsymbol{J}}{P}-\frac{T^{2}D_{a}}{\gamma}\int_{0}^{\tau}dt\int d\mathbf{r}\frac{(\nabla P)^{2}}{P}+\frac{3TD_{a}\tau}{\tau_{p}}\nonumber \\
 & = & TD_{a}\Delta S-\frac{T^{2}D_{a}}{\gamma}\int_{0}^{\tau}dt\int d\mathbf{r}\frac{(\nabla P)^{2}}{P}+\frac{3TD_{a}\tau}{\tau_{p}},\label{seq:activedotr}
\end{eqnarray}
where we have neglected the second order of $\tau_{p}$ in the seventh
equality. In the second equality, we have the following considerations:
The property of the colored noise $\boldsymbol{\eta}(t)$ indicates
that
\begin{eqnarray}
\langle\eta_{i}(t)\eta_{j}(t')\rangle & = & \text{\ensuremath{\frac{\gamma TD_{a}}{\tau_{p}}}\ensuremath{\delta_{ij}}}\mathrm{e}^{-|t-t'|/\tau_{p}}\nonumber \\
 & \stackrel{\tau_{p}\to0}{=} & 2\gamma TD_{a}\ensuremath{\delta_{ij}}\delta(t-t').\label{seq:varcolored}
\end{eqnarray}
However, when we consider the condition $t=t'$, the variance becomes
\begin{equation}
\langle\eta_{i}(t)\eta_{j}(t)\rangle=\text{\ensuremath{\frac{\gamma TD_{a}}{\tau_{p}}}\ensuremath{\delta_{ij}}}=\frac{\gamma^{2}v^{2}}{3}\delta_{ij}\label{seq:equaltimet}
\end{equation}
which holds as $\tau_{p}\to0$. In the three-dimensional space, we
have
\begin{equation}
\langle\boldsymbol{\eta}(t)\circ\boldsymbol{\eta}(t)\rangle\stackrel{\tau_{p}\to0}{=}\langle\boldsymbol{\xi}_{a}(t)\circ\boldsymbol{\xi}_{a}(t)\rangle=\frac{3\gamma TD_{a}}{\tau_{p}}.\label{seq:etavarlim}
\end{equation}
By using the relations in Eqs.$\ $(\ref{seq:pocurrent}) and$\ $(\ref{seq:activedotr}),
we find the mean absorbed heat as:
\begin{eqnarray}
Q & = & T\Delta S-\frac{3TD_{a}\tau}{\tau_{p}}-\gamma\int_{0}^{\tau}dt\int d\mathbf{r}\frac{\boldsymbol{J}^{2}}{P}+\frac{T^{2}D_{a}}{\gamma}\int_{0}^{\tau}dt\int d\mathbf{r}\frac{(\nabla P)^{2}}{P}.\label{seq:meanheatasb}
\end{eqnarray}
Considering the activity parameter $D_{a}\equiv\gamma v^{2}\tau_{p}/(3T)$
in Eq.$\ $(\ref{seq:meanheatasb}), we finally obtain:
\begin{equation}
Q=T\Delta S-\gamma v^{2}\tau-\int_{0}^{\tau}dt\int d\mathbf{r}[\frac{\gamma\boldsymbol{J}^{2}}{P}-\frac{T\tau_{p}v^{2}(\nabla P)^{2}}{3P}].\label{seq:maintextheat}
\end{equation}
This result is equivalent to Eq.$\ $(6) in the main text.

\section{The optimal control of active matter\label{SSec-five}}

In this section, the optimization of the control protocols for active
matter is derived using two approaches: optimal transport theory and
the thermodynamic geometry scheme.

\subsection{Optimal transport theory\label{SSec-fiveA}}

The primary objective is to minimize the irreversible part of the
mean work, $W_{o}=\int_{0}^{\tau}dt\int d\mathbf{r}\gamma\boldsymbol{J}^{2}/P$.
To achieve this, the system must satisfy the constraint given by the
Fokker-Planck equation:
\begin{equation}
\frac{\partial P(\mathbf{r},t)}{\partial t}=-\nabla\cdot\boldsymbol{J}(\mathbf{r},t)=-\nabla\cdot(\mathbf{u}(\mathbf{r},t)P(\mathbf{r},t)),\label{seq:fpequati}
\end{equation}
where $\mathbf{u}(\mathbf{r},t)$ is the current velocity, defined
as:
\begin{equation}
\mathbf{u}(\mathbf{r},t)\equiv\frac{\boldsymbol{J}(\mathbf{r},t)}{P(\mathbf{r},t)}=-\frac{1}{\gamma}[\nabla U(\mathbf{r},t)+T(1+D_{a})\nabla\ln P(\mathbf{r},t)].\label{seq:currentspeed}
\end{equation}
To optimize $W_{o}$, the Lagrangian multiplier method is used, incorporating
the constraint in the Fokker-Planck equation \cite{Dechant2019}.
The Lagrangian is expressed as:
\begin{eqnarray}
\mathcal{L} & = & \int_{0}^{\tau}dt\int d\mathbf{r}\{\frac{\gamma(\boldsymbol{J}(\mathbf{r},t))^{2}}{P(\mathbf{r},t)}-2\phi(\mathbf{r},t)[\frac{\partial P(\mathbf{r},t)}{\partial t}+\nabla\cdot(\mathbf{u}(\mathbf{r},t)P(\mathbf{r},t))]+2\psi(\mathbf{r},t)[\mathbf{u}(\mathbf{r},t)+\frac{1}{\gamma}\nabla U(\mathbf{r},t)\nonumber \\
 &  & +\frac{T(1+D_{a})}{\gamma}\nabla\ln P(\mathbf{r},t)]\}\nonumber \\
 & = & \int_{0}^{\tau}dt\int d\mathbf{r}\{\gamma(\mathbf{u}(\mathbf{r},t))^{2}P(\mathbf{r},t)-2\phi(\mathbf{r},t)[\frac{\partial P(\mathbf{r},t)}{\partial t}+\nabla\cdot(\mathbf{u}(\mathbf{r},t)P(\mathbf{r},t))]+2\psi(\mathbf{r},t)[\mathbf{u}(\mathbf{r},t)+\frac{1}{\gamma}\nabla U(\mathbf{r},t)\nonumber \\
 &  & +\frac{T(1+D_{a})}{\gamma}\nabla\ln P(\mathbf{r},t)]\},\label{seq:lagrangmul}
\end{eqnarray}
where $\phi(\mathbf{r},t)$ and $\psi(\mathbf{r},t)$ represent two
Lagrangian multipliers. By optimizing over $\mathbf{u}(\mathbf{r},t)$,
$P(\mathbf{r},t)$, $\phi(\mathbf{r},t)$, $\psi(\mathbf{r},t)$,
and $U(\mathbf{r},t)$, the following equations are derived:
\begin{eqnarray}
\frac{\delta\mathcal{L}}{\delta\mathbf{u}} & = & 2\gamma\mathbf{u}(\mathbf{r},t)P(\mathbf{r},t)+2(\nabla\phi(\mathbf{r},t))P(\mathbf{r},t)+2\psi(\mathbf{r},t)=0,\nonumber \\
\frac{\delta\mathcal{L}}{\delta P} & = & \gamma(\mathbf{u}(\mathbf{r},t))^{2}+2\frac{\partial\phi(\mathbf{r},t)}{\partial t}+2(\nabla\phi(\mathbf{r},t))\mathbf{u}(\mathbf{r},t)-\frac{2T(1+D_{a})}{\gamma}\frac{\nabla\psi(\mathbf{r},t)}{P(\mathbf{r},t)}=0,\nonumber \\
\frac{\delta\mathcal{L}}{\delta\phi} & = & \frac{\partial P(\mathbf{r},t)}{\partial t}+\nabla\cdot(\mathbf{u}(\mathbf{r},t)P(\mathbf{r},t))=0,\nonumber \\
\frac{\delta\mathcal{L}}{\delta\psi} & = & \mathbf{u}(\mathbf{r},t)+\frac{1}{\gamma}\nabla U(\mathbf{r},t)+\frac{T(1+D_{a})}{\gamma}\nabla\ln P(\mathbf{r},t)=0,\nonumber \\
\frac{\delta\mathcal{L}}{\delta U} & = & -\frac{2}{\gamma}\nabla\psi(\mathbf{r},t)=0.\label{seq:mutiresu}
\end{eqnarray}
These simplify into:
\begin{eqnarray}
\psi(\mathbf{r},t) & = & 0,\label{seq:phide0}\\
\mathbf{u}(\mathbf{r},t) & = & -\frac{1}{\gamma}\nabla\phi(\mathbf{r},t),\label{seq:unabphi}\\
\nabla U(\mathbf{r},t) & = & \nabla\phi(\mathbf{r},t)-T(1+D_{a})\nabla\ln P(\mathbf{r},t),\label{seq:nablauphi}\\
\frac{\partial P(\mathbf{r},t)}{\partial t} & = & -\nabla\cdot(\mathbf{u}(\mathbf{r},t)P(\mathbf{r},t)),\label{seq:pnabup}\\
\frac{\partial\phi(\mathbf{r},t)}{\partial t} & = & \frac{1}{2\gamma}(\nabla\phi(\mathbf{r},t))^{2}.\label{seq:phinabphi}
\end{eqnarray}
Normalizing time as $s\equiv t/\tau$ in Eq.$\ $(\ref{seq:phinabphi}),
we obtain the scaling relations $\phi(\mathbf{r},t)=\Phi(\mathbf{r},s)/\tau$
and $\mathbf{u}(\mathbf{r},t)=\boldsymbol{u}(\mathbf{r},s)/\tau$,
which lead to $P(\mathbf{r},t)=\mathcal{P}(\mathbf{r},s)$ and $\boldsymbol{J}(\mathbf{r},t)=\mathcal{\boldsymbol{J}}(\mathbf{r},s)/\tau$.
These indicate that driven by the optimal control protocol solved
from optimal transport theory, the irreversible part of the mean work
follows a scaling relation as
\begin{eqnarray}
W_{o} & = & \int_{0}^{\tau}dt\int d\mathbf{r}\frac{\gamma(\boldsymbol{J}(\mathbf{r},t))^{2}}{P(\mathbf{r},t)}\nonumber \\
 & = & \frac{1}{\tau}\int_{0}^{1}ds\int d\mathbf{r}\frac{\gamma(\mathcal{\boldsymbol{J}}(\mathbf{r},s))^{2}}{\mathcal{P}(\mathbf{r},s)}.\label{seq:scalingirrwork}
\end{eqnarray}
Similarly, the entropy production induced by the optimal control follows
the scaling relation:
\begin{eqnarray}
\Delta S_{\mathrm{tot}} & = & \frac{\gamma v^{2}\tau}{T}+\int_{0}^{\tau}dt\int d\mathbf{r}[\frac{\gamma\boldsymbol{J}^{2}}{TP}-\frac{\tau_{p}v^{2}(\nabla P)^{2}}{3P}]\nonumber \\
 & = & \frac{1}{\tau}\int_{0}^{1}ds\int d\mathbf{r}\frac{\gamma(\mathcal{\boldsymbol{J}}(\mathbf{r},s))^{2}}{T\mathcal{P}(\mathbf{r},s)}+\tau\frac{\gamma v^{2}}{T}[1-\tau_{p}\int_{0}^{1}ds\int d\mathbf{r}\frac{T(\nabla\mathcal{P}(\mathbf{r},s))^{2}}{3\gamma\mathcal{P}(\mathbf{r},s)}]\nonumber \\
 & = & \frac{A_{o}}{T}\frac{1}{\tau}+\frac{E_{a}}{T}\tau,\label{eq:scalingentropro}
\end{eqnarray}
where $A_{o}\equiv\gamma\int_{0}^{1}ds\int d\mathbf{r}\mathcal{\boldsymbol{J}}^{2}/\mathcal{P}$
and $E_{a}\equiv\gamma v^{2}[1-\tau_{p}\int_{0}^{1}ds\int d\mathbf{r}T(\nabla\mathcal{P})^{2}/(3\gamma\mathcal{P})]$.
This result is equivalent to the one in Eq.$\ $(8) of the main text.

\subsection{The thermodynamic geometry scheme\label{SSec-fiveB}}

In the framework of thermodynamic geometry, the irreversible part
of the mean work is obtained in Eq.$\ $(\ref{seq:irrpartwork}) as
$W_{o}=\sum_{\mu\nu}\int_{0}^{\tau}dt\dot{\lambda}_{\mu}\dot{\lambda}_{\nu}g_{\mu\nu}$.
The optimal control protocol is found by solving the geodesic equation:
\begin{equation}
\ddot{\lambda}_{\mu}+\sum_{\nu\kappa}\Gamma_{\nu\kappa}^{\mu}\dot{\lambda}_{\nu}\dot{\lambda}_{\kappa}=0,\label{seq:geodesicequt}
\end{equation}
where the Christoffel symbol is defined as $\Gamma_{\nu\kappa}^{\mu}\equiv(1/2)\sum_{\iota}(g^{-1})_{\iota\mu}(\partial_{\lambda_{\kappa}}g_{\iota\nu}+\partial_{\lambda_{\nu}}g_{\iota\kappa}-\partial_{\lambda_{\iota}}g_{\nu\kappa})$.
By normalizing time as $s\equiv t/\tau$ in Eq.$\ $(\ref{seq:geodesicequt}),
we arrive at the scaling relation of the optimal control as $\boldsymbol{\lambda}(t)=\boldsymbol{\lambda}(s)$,
which leads to $P(\mathbf{r},\boldsymbol{\lambda}(t))=\mathcal{P}(\mathbf{r},\boldsymbol{\lambda}(s))$
and $\boldsymbol{J}(\mathbf{r},t)=\mathcal{\boldsymbol{J}}(\mathbf{r},s)/\tau$.
These conclusions lead to the same scaling relations as obtained from
the optimal transport theory, i.e., Eqs.$\ $(\ref{seq:scalingirrwork})
and$\ $(\ref{eq:scalingentropro}).

\section{The performance of an active monothermal engine\label{SSec-teen}}

\subsection{The maximum efficiency\label{SSec-twoA}}

The efficiency of the active monothermal engine is defined as:
\begin{equation}
\varepsilon\equiv\frac{W_{\mathrm{tot}}}{W_{\mathrm{tot}}+Q_{\mathrm{tot}}}=\frac{T(D_{a}^{h}-D_{a}^{l})\Delta S-A_{o}/\tau_{h}-A_{o}/\tau_{l}}{T(D_{a}^{h}-D_{a}^{l})\Delta S+E_{a}^{h}\tau_{h}+E_{a}^{l}\tau_{l}}.\label{seq:activeefficiency}
\end{equation}
The scaling irreversible work $A_{o}=\gamma\mathcal{W}^{2}$ is uniform
for both isoactive processes due to the symmetry of the Wasserstein
distance $\mathcal{W}(P_{a},P_{b})=\mathcal{W}(P_{b},P_{a})$. To
optimize this efficiency with respect to the control durations $\tau_{h}$
and $\tau_{l}$, we derive the following equations:
\begin{eqnarray}
[T(D_{a}^{h}-D_{a}^{l})\Delta S+E_{a}^{h}\tau_{h}+E_{a}^{l}\tau_{l}]A_{o}/\tau_{h}^{2} & = & [T(D_{a}^{h}-D_{a}^{l})\Delta S-A_{o}/\tau_{h}-A_{o}/\tau_{l}]E_{a}^{h},\nonumber \\{}
[T(D_{a}^{h}-D_{a}^{l})\Delta S+E_{a}^{h}\tau_{h}+E_{a}^{l}\tau_{l}]A_{o}/\tau_{l}^{2} & = & [T(D_{a}^{h}-D_{a}^{l})\Delta S-A_{o}/\tau_{h}-A_{o}/\tau_{l}]E_{a}^{l}.\label{seq:maxeffequas}
\end{eqnarray}
Solving these yields the optimal durations for the high activity stage
and the low activity stage:
\begin{eqnarray}
\tau_{h}^{e} & = & \frac{A_{o}(1+\sqrt{E_{a}^{l}/E_{a}^{h}})}{T(D_{a}^{h}-D_{a}^{l})\Delta S}+\sqrt{\frac{A_{o}^{2}(1+\sqrt{E_{a}^{l}/E_{a}^{h}})^{2}}{T^{2}(D_{a}^{h}-D_{a}^{l})^{2}(\Delta S)^{2}}+\frac{A_{o}}{E_{a}^{h}}},\nonumber \\
\tau_{l}^{e} & = & \frac{A_{o}(1+\sqrt{E_{a}^{h}/E_{a}^{l}})}{T(D_{a}^{h}-D_{a}^{l})\Delta S}+\sqrt{\frac{A_{o}^{2}(1+\sqrt{E_{a}^{h}/E_{a}^{l}})^{2}}{T^{2}(D_{a}^{h}-D_{a}^{l})^{2}(\Delta S)^{2}}+\frac{A_{o}}{E_{a}^{l}}}.\label{seq:optefftim}
\end{eqnarray}
The maximum efficiency is then:
\begin{equation}
\varepsilon_{\mathrm{max}}=(\frac{\tau_{h}^{*}}{\tau_{h}^{e}})^{2}=(\frac{\tau_{l}^{*}}{\tau_{l}^{e}})^{2},\label{seq:maxeffiency}
\end{equation}
where $\tau_{h}^{*}\equiv\sqrt{A_{o}/E_{a}^{h}}$ and $\tau_{l}^{*}\equiv\sqrt{A_{o}/E_{a}^{l}}$
represent the optimal driving durations for minimum entropy production
in the two isoactive processes.

Comparing both sides of Eq.$\ $(\ref{seq:phinabphi}), we conclude
that $\phi(\mathbf{r},t)$ is independent of the activity $v$, which
leads to the independence of $v$ for the optimal control of $P(\mathbf{r},t)$
and $\boldsymbol{J}(\mathbf{r},t)$. Then, we finally derive that
$A_{o}\equiv\gamma\int_{0}^{1}ds\int d\mathbf{r}\mathcal{\boldsymbol{J}}^{2}/\mathcal{P}$
is independent of $v$ and $E_{a}\equiv\gamma v^{2}[1-\tau_{p}\int_{0}^{1}ds\int d\mathbf{r}T(\nabla\mathcal{P})^{2}/(3\gamma\mathcal{P})]=F_{a}v^{2}$
is proportional to $v^{2}$ with the function $F_{a}$ independent
of $v$. These results lead to

\begin{eqnarray}
\frac{\tau_{h}^{e}}{\tau_{h}^{*}}=\frac{\tau_{l}^{e}}{\tau_{l}^{*}} & = & \frac{\sqrt{A_{o}}(\sqrt{E_{a}^{h}}+\sqrt{E_{a}^{l}})}{T(D_{a}^{h}-D_{a}^{l})\Delta S}+\sqrt{\frac{A_{o}(\sqrt{E_{a}^{h}}+\sqrt{E_{a}^{l}})^{2}}{T^{2}(D_{a}^{h}-D_{a}^{l})^{2}(\Delta S)^{2}}+1}\nonumber \\
 & = & \frac{\sqrt{A_{o}}(\sqrt{F_{a}^{h}}v_{h}+\sqrt{F_{a}^{l}}v_{l})}{(\gamma\tau_{p}\Delta S/3)(v_{h}^{2}-v_{l}^{2})}+\sqrt{\frac{A_{o}(\sqrt{F_{a}^{h}}v_{h}+\sqrt{F_{a}^{l}}v_{l})^{2}}{(\gamma\tau_{p}\Delta S/3)^{2}(v_{h}^{2}-v_{l}^{2})^{2}}+1}.\label{seq:scalingtauras}
\end{eqnarray}
By fixing the lower activity $v_{l}$, we obtain the function relationship
between the maximum efficiency $\varepsilon_{\mathrm{max}}$ and the
ratio between activity levels $v_{h}/v_{l}$, which is drawn in Fig.$\ $2(b)
of the main text. In the limit $v_{h}/v_{l}\to1$, the maximum efficiency
satisfies the relation $\varepsilon_{\mathrm{max}}=\alpha_{1}(v_{h}/v_{l}-1)^{2}$
where $\alpha_{1}\equiv4v_{l}^{2}\gamma^{2}\tau_{p}^{2}(\Delta S)^{2}/[9A_{o}(\sqrt{F_{a}^{h}}+\sqrt{F_{a}^{l}})^{2}]$.
While in the large ratio limit $v_{h}/v_{l}\to\infty$, the maximum
efficiency yields $\varepsilon_{\mathrm{max}}=1-\alpha_{2}v_{h}/v_{l}$
where $\alpha_{2}=9\sqrt{A_{o}F_{a}^{h}}/(v_{l}\gamma\tau_{p}\Delta S)$.

\subsection{The maximum power\label{SSec-twoB}}

The power of the active monothermal engine follows as
\begin{equation}
\mathrm{P}\equiv\frac{-W_{\mathrm{tot}}}{\tau_{h}+\tau_{l}}=\frac{T(D_{a}^{h}-D_{a}^{l})\Delta S-A_{o}/\tau_{h}-A_{o}/\tau_{l}}{\tau_{h}+\tau_{l}}.\label{seq:definepower}
\end{equation}
By taking derivative over the durations $\tau_{h}$ and $\tau_{l}$,
we obtain:
\begin{eqnarray}
2A_{o}/\tau_{h}+A_{o}\tau_{l}/\tau_{h}^{2}+A_{o}/\tau_{l} & = & T(D_{a}^{h}-D_{a}^{l})\Delta S,\nonumber \\
2A_{o}/\tau_{l}+A_{o}\tau_{h}/\tau_{l}^{2}+A_{o}/\tau_{h} & = & T(D_{a}^{h}-D_{a}^{l})\Delta S.\label{seq:maxpoequs}
\end{eqnarray}
Solving these equations gives the durations for maximum power:
\begin{equation}
\tau_{h}^{p}=\tau_{l}^{p}=\frac{4A_{o}}{T(D_{a}^{h}-D_{a}^{l})\Delta S}.\label{seq:maxpowtau}
\end{equation}
The maximum power is:
\begin{equation}
\mathrm{P}_{\mathrm{max}}=\frac{T^{2}(D_{a}^{h}-D_{a}^{l})^{2}(\Delta S)^{2}}{16A_{o}},\label{seq:maxpowere}
\end{equation}
which is proportional to $[(v_{h}/v_{l})^{2}-1]^{2}$ when the lower
activity $v_{l}$ is fixed. This result is represented in Fig.$\ $2(b)
of the main text. The efficiency at maximum power is then obtained
as
\begin{eqnarray}
\varepsilon_{p} & = & \frac{1}{2\{1+A_{o}(E_{a}^{h}+E_{a}^{l})/[T^{2}(D_{a}^{h}-D_{a}^{l})^{2}(\Delta S)^{2}]\}}\nonumber \\
 & = & \frac{1}{2\{1+A_{o}(F_{a}^{h}v_{h}^{2}+F_{a}^{l}v_{l}^{2})/[(\gamma\tau_{p}\Delta S/3)^{2}(v_{h}^{2}-v_{l}^{2})^{2}]\}},\label{seq:effmaxpow}
\end{eqnarray}
which approaches $1/2$ in the large ratio limit $v_{h}/v_{l}\to\infty$.

\bibliographystyle{apsrev4-1}
\bibliography{ref}